\documentclass[12pt,preprint]{aastex}
\usepackage{amsmath}
\usepackage{float}
\title{Discovery of very-high-energy emission from RGB J2243+203 and derivation of its redshift upper limit}

\author{
A.~U.~Abeysekara\altaffilmark{1},
S.~Archambault\altaffilmark{2},
A.~Archer\altaffilmark{3},
W.~Benbow\altaffilmark{4},
R.~Bird\altaffilmark{5},
R.~Brose\altaffilmark{6,7},
M.~Buchovecky\altaffilmark{5},
J.~H.~Buckley\altaffilmark{3},
V.~Bugaev\altaffilmark{3},
M.~Cerruti\altaffilmark{4},
M.~P.~Connolly\altaffilmark{8},
W.~Cui\altaffilmark{9,10},
A.~Falcone\altaffilmark{11},
Q.~Feng\altaffilmark{2},
J.~P.~Finley\altaffilmark{9},
H.~Fleischhack\altaffilmark{7},
L.~Fortson\altaffilmark{12},
A.~Furniss\altaffilmark{13},
G.~H.~Gillanders\altaffilmark{8},
S.~Griffin\altaffilmark{2},
J.~Grube\altaffilmark{14},
M.~H\"utten\altaffilmark{7},
D.~Hanna\altaffilmark{2},
O.~Hervet\altaffilmark{15},
J.~Holder\altaffilmark{16},
T.~B.~Humensky\altaffilmark{17},
C.~A.~Johnson\altaffilmark{15},
P.~Kaaret\altaffilmark{18},
P.~Kar\altaffilmark{1},
N.~Kelley-Hoskins\altaffilmark{7},
M.~Kertzman\altaffilmark{19},
D.~Kieda\altaffilmark{1},
M.~Krause\altaffilmark{7},
F.~Krennrich\altaffilmark{20},
S.~Kumar\altaffilmark{16},
M.~J.~Lang\altaffilmark{8},
G.~Maier\altaffilmark{7},
S.~McArthur\altaffilmark{9},
P.~Moriarty\altaffilmark{8},
R.~Mukherjee\altaffilmark{21},
D.~Nieto\altaffilmark{17},
S.~O'Brien\altaffilmark{22},
R.~A.~Ong\altaffilmark{5},
A.~N.~Otte\altaffilmark{23},
N.~Park\altaffilmark{24},
A.~Petrashyk\altaffilmark{17},
M.~Pohl\altaffilmark{6,7},
A.~Popkow\altaffilmark{5},
E.~Pueschel\altaffilmark{7},
J.~Quinn\altaffilmark{22},
K.~Ragan\altaffilmark{2},
P.~T.~Reynolds\altaffilmark{25},
G.~T.~Richards\altaffilmark{23},
E.~Roache\altaffilmark{4},
C.~Rulten\altaffilmark{12},
I.~Sadeh\altaffilmark{7},
M.~Santander\altaffilmark{21},
G.~H.~Sembroski\altaffilmark{9},
K.~Shahinyan\altaffilmark{12},
D.~Staszak\altaffilmark{24},
I.~Telezhinsky\altaffilmark{6,7},
J.~Tyler\altaffilmark{2},
V.~V.~Vassiliev\altaffilmark{5},
S.~P.~Wakely\altaffilmark{24},
O.~M.~Weiner\altaffilmark{17},
A.~Weinstein\altaffilmark{20},
P.~Wilcox\altaffilmark{18},
A.~Wilhelm\altaffilmark{6,7},
D.~A.~Williams\altaffilmark{15},
B.~Zitzer\altaffilmark{2}
}

\altaffiltext{1}{Department of Physics and Astronomy, University of Utah, Salt Lake City, UT 84112, USA}
\altaffiltext{2}{Physics Department, McGill University, Montreal, QC H3A 2T8, Canada}
\altaffiltext{3}{Department of Physics, Washington University, St. Louis, MO 63130, USA}
\altaffiltext{4}{Fred Lawrence Whipple Observatory, Harvard-Smithsonian Center for Astrophysics, Amado, AZ 85645, USA}
\altaffiltext{5}{Department of Physics and Astronomy, University of California, Los Angeles, CA 90095, USA}
\altaffiltext{6}{Institute of Physics and Astronomy, University of Potsdam, 14476 Potsdam-Golm, Germany}
\altaffiltext{7}{DESY, Platanenallee 6, 15738 Zeuthen, Germany}
\altaffiltext{8}{School of Physics, National University of Ireland Galway, University Road, Galway, Ireland}
\altaffiltext{9}{Department of Physics and Astronomy, Purdue University, West Lafayette, IN 47907, USA}
\altaffiltext{10}{Department of Physics and Center for Astrophysics, Tsinghua University, Beijing 100084, China.}
\altaffiltext{11}{Department of Astronomy and Astrophysics, 525 Davey Lab, Pennsylvania State University, University Park, PA 16802, USA}
\altaffiltext{12}{School of Physics and Astronomy, University of Minnesota, Minneapolis, MN 55455, USA}
\altaffiltext{13}{Department of Physics, California State University - East Bay, Hayward, CA 94542, USA}
\altaffiltext{14}{Department of Physics, Stevens Institute of Technology, Hoboken, NJ 07030, USA}
\altaffiltext{15}{Santa Cruz Institute for Particle Physics and Department of Physics, University of California, Santa Cruz, CA 95064, USA}
\altaffiltext{16}{Department of Physics and Astronomy and the Bartol Research Institute, University of Delaware, Newark, DE 19716, USA}
\altaffiltext{17}{Physics Department, Columbia University, New York, NY 10027, USA}
\altaffiltext{18}{Department of Physics and Astronomy, University of Iowa, Van Allen Hall, Iowa City, IA 52242, USA}
\altaffiltext{19}{Department of Physics and Astronomy, DePauw University, Greencastle, IN 46135-0037, USA}
\altaffiltext{20}{Department of Physics and Astronomy, Iowa State University, Ames, IA 50011, USA}
\altaffiltext{21}{Department of Physics and Astronomy, Barnard College, Columbia University, NY 10027, USA}
\altaffiltext{22}{School of Physics, University College Dublin, Belfield, Dublin 4, Ireland}
\altaffiltext{23}{School of Physics and Center for Relativistic Astrophysics, Georgia Institute of Technology, 837 State Street NW, Atlanta, GA 30332-0430}
\altaffiltext{24}{Enrico Fermi Institute, University of Chicago, Chicago, IL 60637, USA}
\altaffiltext{25}{Department of Physical Sciences, Cork Institute of Technology, Bishopstown, Cork, Ireland}

\begin{document}


\begin{abstract}
Very-high-energy (VHE; $>$ 100 GeV) gamma-ray emission from the blazar RGB J2243+203 was discovered with the VERITAS Cherenkov telescope array, during the period between 21 and 24 December 2014.
The VERITAS energy spectrum from this source can be fit by a power law with a photon index of $4.6 \pm 0.5$, and a flux normalization at 0.15 TeV of $(6.3 \pm 1.1) \times 10^{-10} ~ \textrm{cm}^{-2} \textrm{s}^{-1} \textrm{TeV}^{-1}$.
The integrated \textit{Fermi}-LAT flux from 1 GeV to 100 GeV during the VERITAS detection is $(4.1 \pm 0.8) \times 10^{\textrm{-8}} ~\textrm{cm}^{\textrm{-2}}\textrm{s}^{\textrm{-1}}$, which is an order of magnitude larger than the four-year-averaged flux in the same energy range reported in the 3FGL catalog, ($4.0 \pm 0.1 \times 10^{\textrm{-9}} ~ \textrm{cm}^{\textrm{-2}}\textrm{s}^{\textrm{-1}}$).
The detection with VERITAS triggered observations in the X-ray band with the \textit{Swift}-XRT.
However, due to scheduling constraints \textit{Swift}-XRT observations were performed 67 hours after the VERITAS detection, not simultaneous with the VERITAS observations.
The observed X-ray energy spectrum between 2 keV and 10 keV can be fitted with a power-law with a spectral index of $2.7 \pm 0.2$, 
and the integrated photon flux in the same energy band is $(3.6 \pm 0.6) \times 10^{-13} ~\textrm{cm}^{-2} \textrm{s}^{-1}$.
EBL model-dependent upper limits of the blazar redshift have been derived.
Depending on the EBL model used, the upper limit varies in the range from z $<~0.9$ to z $<~1.1$.
\end{abstract}

\section{Introduction}\label{Sec:Introduction}
RGB J2243+203 was first reported as a radio source in the MIT-Green Bank (RGB) catalog by \cite{Griffith1990}.
The optical energy spectrum of the source was studied by \cite{Laurent-Muehleisen98} who found the 
observed energy spectrum to be featureless, leading its classifications as a BL Lac type blazar.
\cite{Laurent-Muehleisen} studied the cross-correlation of a ROSAT All-Sky Survey 
source list with an extensive sample of radio sources from a RGB sky survey, including RGB J2243+203, 
and found the X-ray to radio flux ratios of the RGB BL Lacertae objects are shown to peak between 
the low-energy-peaked BL Lacs (LBLs) and the high-energy-peaked BL Lacs (HBLs). 
Therefore, \cite{Laurent-Muehleisen} categorized RGB J2243+203 as an intermediate-frequency-peaked BL Lac (IBL).
Twelve years later, the second \textit{Fermi}-LAT AGN catalog (2LAC) \citep{Ackermann2011}  identified RGB J2243+203 as a BL Lac  object having a synchrotron peak corresponding to an HBL.
The biggest difference between these two studies is \cite{Laurent-Muehleisen} used the collective properties of a sample of BL Lacertae objects and classified the whole sample as IBLs, 
but \cite{Ackermann2011} used the properties of RGB J2243+203 to classify the source as an HBL.

\cite{Meisner2010} attempted to measure the redshift of RGB J2243+203.
Following \cite{Sbarufatti2005}, \cite{Meisner2010} assumed BL Lac host galaxies are standard candles with $M_R = -22.9 \pm 0.5$,
and estimated the redshift using the apparent magnitude of the host galaxy.
They were unable to resolve the host galaxy of RGB J2243+203.
Therefore, a lower limit on the redshift, $\textrm{z} > 0.39$, was calculated using the upper limit of the host galaxy flux.
However, this method is susceptible to the accuracy of the assumptions.
Section \ref{Redshift} of this paper estimates upper limits for the redshift of the source using the method proposed 
by \cite{Georganopoulos2010}.

The VERITAS discovery of TeV gamma-ray emission from RGB J2243+203 was announced on December 24, 2014 in 
the Astronomer's Telegram \#6849.
Section \ref{VERITASAnalysis} describes the details of the VERITAS results, and Section \ref{FermiAnalysis} describes the analysis of simultaneous and historical \textit{Fermi}-LAT data, and 
Section \ref{SwiftAnalysis} describes the measurements of contemporaneous and historical \textit{Swift}-XRT data.
Section \ref{Redshift} derives EBL model dependent redshift upper limits, and Section \ref{Sec:Discussion} is the discussion.

\section{Observations and Data Analysis}

\subsection{VERITAS observations}\label{VERITASAnalysis}
VERITAS is an array of four imaging atmospheric Cherenkov telescopes (IACTs) that follow the Davies-Cotton design, 
with 12 m diameter reflectors \citep{Holder2008}.
The array is located at the Fred Lawrence Whipple Observatory in southern Arizona ($31^\circ ~ 40^\prime$ N, $110^\circ 57^\prime$ W, 1.3 km
a.s.l.), and is sensitive to gamma rays in the energy range from $\sim 85$ GeV to $\sim30$ TeV.
It has the sensitivity to detect a point source at five standard deviations (at $5 \sigma$) with a brightness of 1\% of the Crab Nebula flux with an exposure of 
$<25$ hours.
The energy of gamma rays can be measured with a resolution of $15-25\%$ and the angular resolution is better than 0.1 degrees at 1TeV.

\begin{figure}[ht]
\centering
 \includegraphics[width=0.5\textwidth]{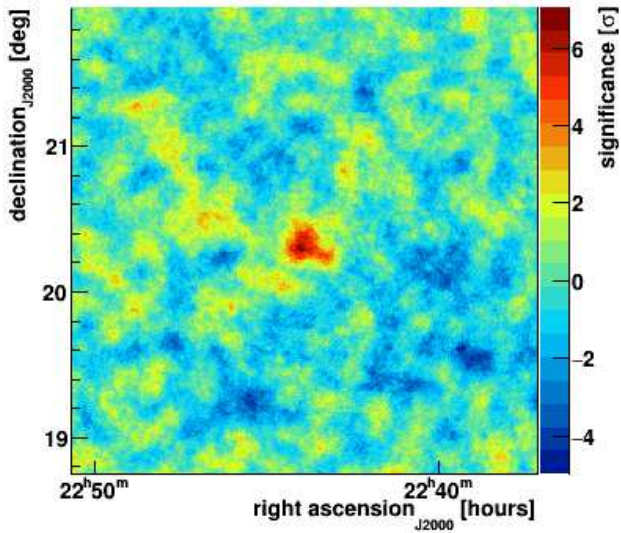}
\caption{VERITAS skymap of the region around RGB J2243+203.}
\label{Fig:VERITASSigmap}
\end{figure}

\begin{figure}
\centering
\includegraphics[width=0.75\textwidth]{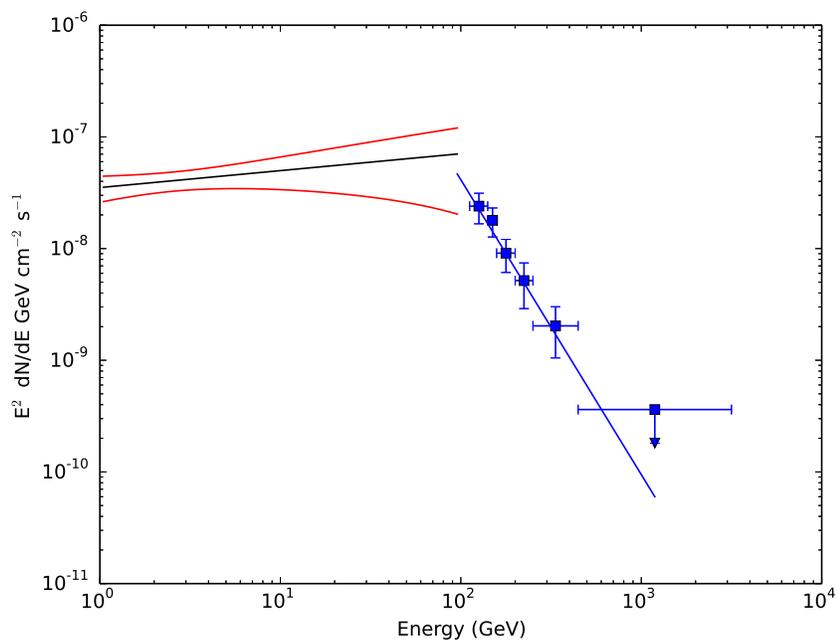}
\caption{Gamma ray SED of RGB J2243+203 during the flare from 2014-12-21 01:44:00 UTC to 2014-12-24 03:34:00 UTC (MJD 57012.072 to MJD 57015.14861).
The solid black line represents the best fit SED of the \textit{Fermi}-LAT detection, and the red bow tie represents the one sigma uncertainty of the likelihood fit to the \textit{Fermi}-LAT data.
VERITAS fluxes are shown as blue points, and the solid blue line represents the best fit to the VERITAS data.}
\label{Fig:MWLSED}
\end{figure}

On 21 December 2014, an automated \textit{Fermi}-LAT analysis pipeline used by VERITAS collaborators indicated that the daily 
\textit{Fermi}-LAT flux of RGB J2243+203 was marginally elevated from the average flux.
VERITAS observed the source from 21 December 2014  (Modified Julian Date (MJD) 57012) through 24 December 2014 (MJD 57015) 
for a total of 264 minutes of good quality data.
Within this time interval, the source was observed from when the night began until the elevation of the source went below $40^\circ$.
The average elevation of the observations is $53^\circ$.
The data set has been analyzed with reflected-region background model \citep{AharonianRefflected}, 
and minimum two telescope triggering criterion described in \cite{Holder2006}.
On the first night, the source was observed for 37 minutes during which 204 ON events and 873 OFF events were detected with a background normalization factor $\alpha$ of 0.167, yielding an excess of 58.5 events corresponding to a detection significance of $4.2 \sigma$.
With an additional 227 minutes, VERITAS detected 1086 ON events, 5393 OFF events with a background normalization factor $\alpha$ of 0.167, 
yielding an increased cumulative excess of 187 events, and the cumulative significance increased to $5.6 \sigma$.
The excess is consistent with a point-like source within the VERITAS point spread function.
Observations ceased on 24 December 2014 because moonlight precluded VERITAS observations with a sufficiently low energy threshold.
The significance map of the region, centered at the source location, is shown in Figure \ref{Fig:VERITASSigmap}.
The gamma-ray energy spectrum above 112 GeV can be fitted with a power-law with spectral index of $4.6 \pm 0.5$, 
and flux normalisation at 0.15 TeV of $(6.3 \pm 1.1) \times 10^{-10} ~\textrm{cm}^{-2} \textrm{s}^{-1} \textrm{TeV}^{-1}$.
The VERITAS gamma-ray energy spectrum is shown in Figure \ref{Fig:MWLSED}.
The best fit has a $\chi^2/NDF$ of 0.55/3 with no significant indication of a spectral break or curvature.
The integrated gamma-ray photon flux above 112 GeV averaged over the VERITAS observations is 
$(7.8 \pm 1.4) \times 10^{-11} ~\textrm{cm}^{-2} \textrm{s}^{-1}$, which corresponds to 14\% of the Crab Nebula flux.

The centroid of the detection was determined by fitting a symmetric 2-dimensional Gaussian to the excess counts map.
The best-fit resulted in a centroid at R.A. = 340.97  (22 h 43 m 52 s) and Dec. = $+20.32^\circ$, 
with a statistical uncertainty of $0.02^\circ$ and systematic uncertainty of $0.007^{\circ}$.
This new very-high-energy source is cataloged as VER J2243+203.
This source position is also consistent with the centroid of the RGB J2243+203 measured using very long baseline interferometry (VLBI), 
with a probability of 0.87.

Before this observation, the blazar was observed by VERITAS between September and October 2009 for a total of about 4 hours of 
good quality data.
The past observations did not show evidence for TeV emission from the blazar, 
and the upper limit on the flux above 170 GeV at the 95\% 
confidence level \citep{ULRolkeAlgo} is $<2.1$\% of the Crab Nebula flux  \citep{VERITASBlazarUL}. 

\begin{figure}[ht]
\centering
 \includegraphics[width=0.8\textwidth]{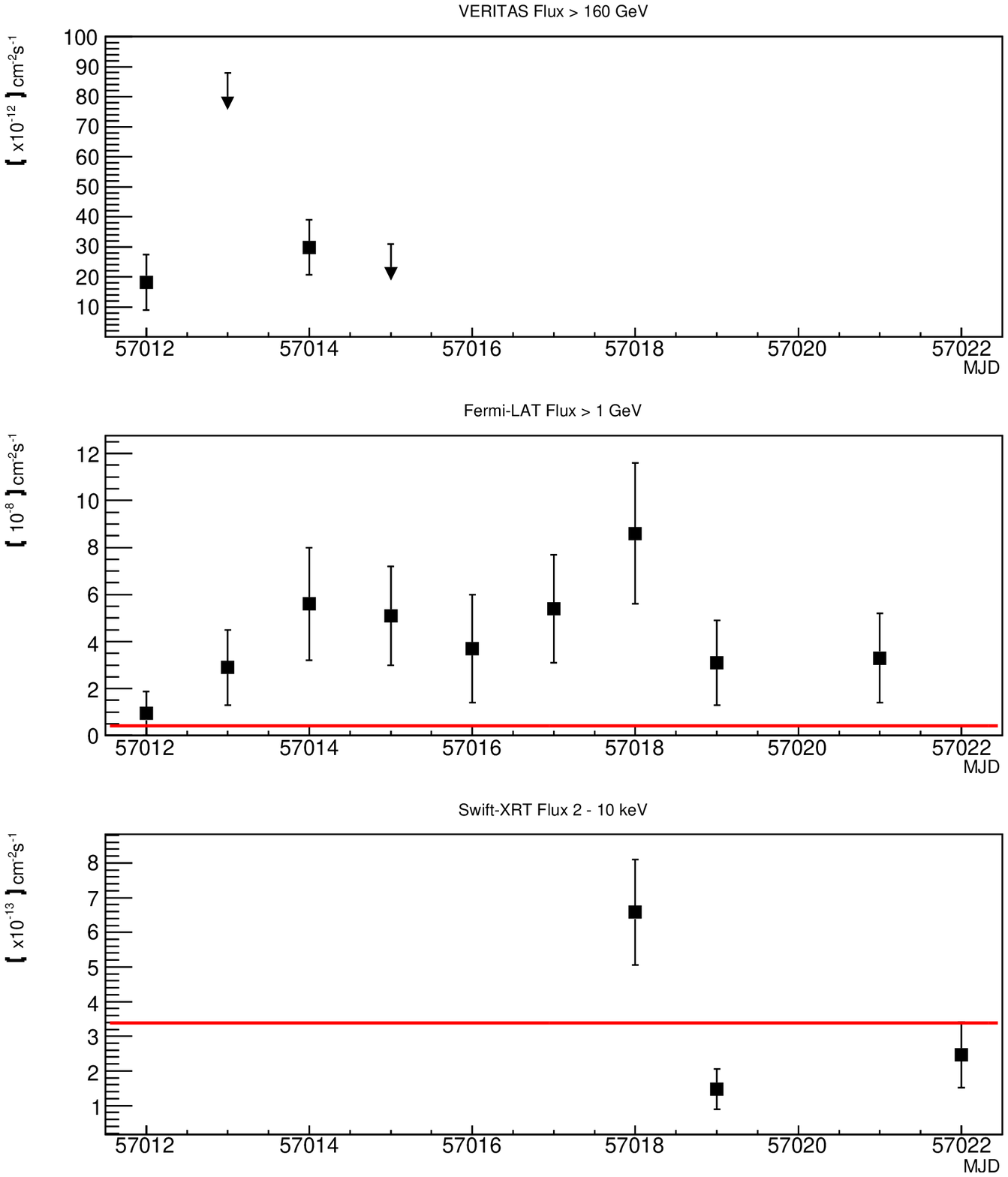}
\caption{\textit{Top Figure}: December 2014 VERITAS light curve from the $21^{\textrm{st}}$ (MJD 57012) to $24^{\textrm{th}}$ (MJD 57015) in nightly bins. 95\% confidence level upper limits are shown for the bins with significance less than 2 $\sigma$.
\textit{Middle Figure}: \textit{Fermi}-LAT light curve from the 21 (MJD 57012) to 31 December 2014 (MJD 57022) in nightly bins. 
The red line shows the four-year average flux as reported in the 3FGL catalog.
\textit{Bottom Figure}: \textit{Swift}-XRT light curve. The red line shows the best fit to a constant.
}
\label{Fig:VERITASLightCurve}
\end{figure}

The VERITAS light curve from 21 December 2014 (MJD 57012) to 24 December 2014 (MJD 57015) in nightly bins is shown 
in the top subfigure in Figure~\ref{Fig:VERITASLightCurve}. 
On 21 and 23 December 2014, gamma-ray like events were detected with a significance of $4.2 \sigma$ in each day.
No significant detection of gamma-ray like events was recorded at the source location on 22 and 24 December 2014.
Therefore, 95\% upper limits are marked for those two days.

\subsection{\textit{Fermi} Large Area Telescope (LAT) observations}\label{FermiAnalysis}

The most recent \textit{Fermi}-LAT source catalog, 3FGL, \citep{Acero2015} reported detection of sources in the energy range of 100 MeV 
to 300 GeV, using the first four years of science data from \textit{Fermi}-LAT.
RGB J2243+203 (3FGL J2243.9+2021) was detected with a statistical significance of $45.9\sigma$, and the average flux between 1 GeV and 100 GeV 
in the 4 year data set is $(4.1 \pm 0.2) \times 10^{-9}$ cm$^{-2}$ s$^{-1}$.
The variability of the flux is measured using the variability index defined in \cite{Abdo1FGL} with an index of 59.1 for the four year data set.
The energy spectrum in the 100 MeV to 100 GeV energy range was described as a simple power-law energy spectrum with spectral index of 
$1.79 \pm 0.03$.

The Second Fermi-LAT Catalog of High-Energy Sources (2FHL) \citep{Ackermann2016} also reported a detection of RGB J2243+203 at energies above 50 GeV.
The source was detected with a $\textrm{TS}~=~123.5$, and the average flux between 50 GeV and 2 TeV is $4.09 \pm 1.24 \times 10^{-11} ~\textrm{cm}^{-2} \textrm{s}^{-1}$.
The energy spectrum was described as a simple power-law spectrum with a spectral index of $4.16 \pm 0.97$.

In order to obtain a quasi-simultaneous GeV gamma-ray energy spectrum of the blazar, the \textit{Fermi}-LAT data set has been analyzed from 
2014-12-21 01:44:00 UTC to 2014-12-24 03:34:00 UTC (MJD 57012.072 to MJD 57015.14861), overlapping the time
of the VERITAS detection of RGB J2243+203.
The data set has been analyzed using the Pass 8 \textit{Fermi}-LAT analysis tools, with the standard
instrument response function P8R2\_SOURCE\_V6 
The Galactic diffuse emission and the isotropic diffuse emission were modeled with gll\_iem\_v06.fits 
and iso\_P8R2\_SOURCE\_V6\_v06.txt, respectively.
Photons belonging to \textit{Fermi}-LAT event class 128 within a region of interest (ROI) $10^{\circ} \times 10^{\circ}$ 
centered on the blazar position were analyzed, with the selection cuts of rocking angle $< 52^{\circ}$ and zenith angle of $< 90^{\circ}$.
An energy cut with a minimum photon energy of 1 GeV, and maximum photon energy of 100 GeV is applied.

\begin{figure}[ht]
\centering
 \includegraphics[width=0.9\textwidth]{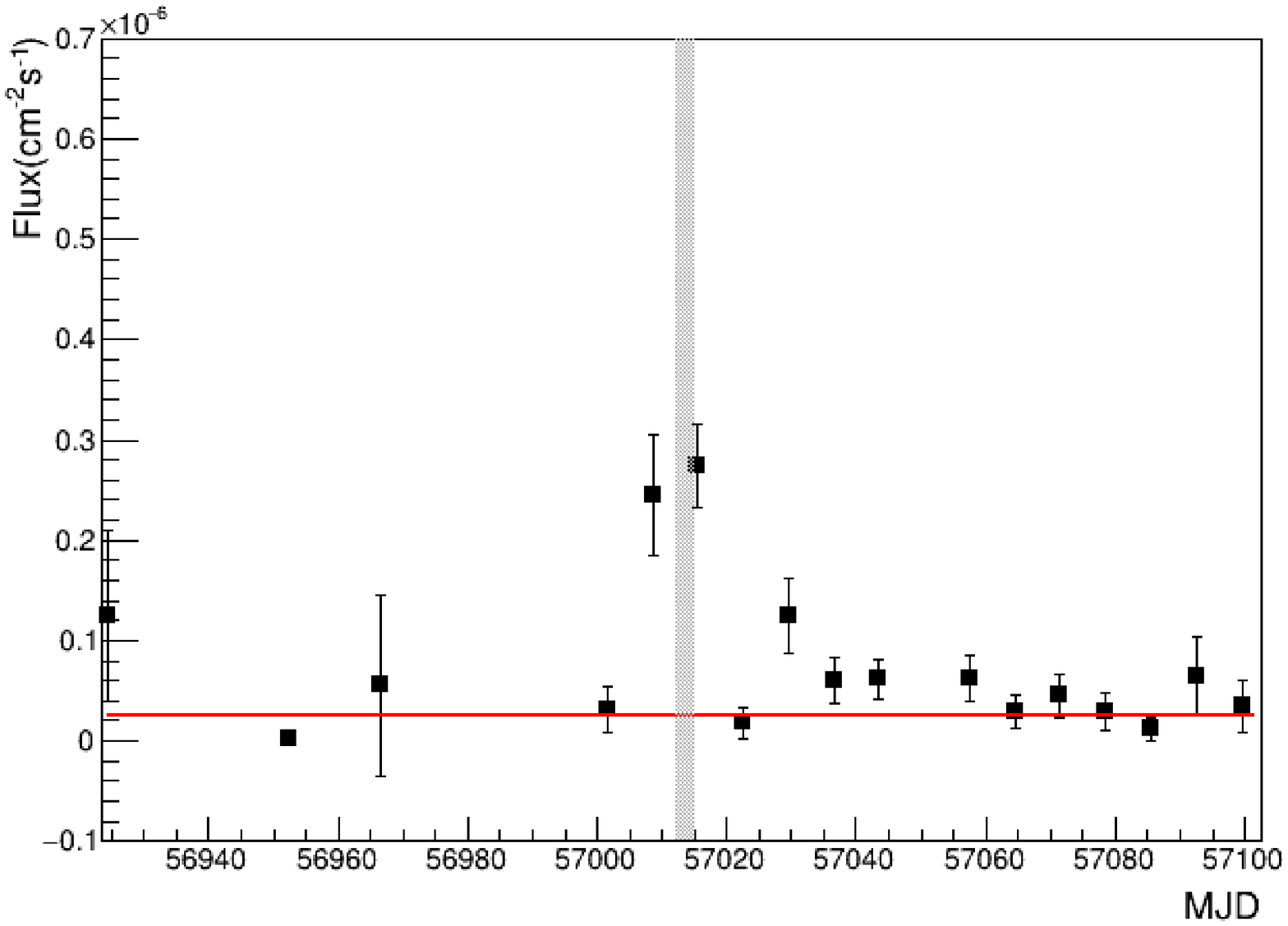}
\caption{\textit{Fermi}-LAT light curve of RGB J2243+203 with seven day bins. 
The solid red line indicates the four-year time-averaged flux reported in the \textit{Fermi}-LAT 3FGL catalog. 
Flux measurements are not plotted for bins with TS less than 4.
The shaded vertical strip is the VERITAS observation window of the source.}
\label{Fig:FermiLongTermLightCurve}
\end{figure}

This analysis yields a detection of RGB J2243+203 with a $\sqrt{TS} = 9.2$.
The gamma-ray energy spectrum between 1 GeV and 100 GeV can be fitted with a power-law with a spectral index of 
$1.8 \pm 0.2$, and a flux normalization at 2.2 GeV of 
$(8.4 \pm 1.7) \times 10^{-12} ~\textrm{cm}^{-2} \textrm{s}^{-1} \textrm{MeV}^{-1}$. 
Spectral shapes more complex than a simple power law do not result in a significant improvement in the fit.
The integrated photon flux averaged over the selected time window is 
$(4.1 \pm 0.8) \times 10^{-8} ~\textrm{cm}^{-2} \textrm{s}^{-1}$, which is an order of magnitude larger than the 
four-year-averaged 3FGL flux in the same energy range; $(4.0 \pm 0.1) \times 10^{-9} ~\textrm{cm}^{-2} \textrm{s}^{-1}$.
The light curve of the integral flux between 1 GeV and 100 GeV with daily binning starting from MJD 57012  
(start of VERITAS observations) to MJD 57022 (the end of \textit{Swift}-XRT observations) is shown in the middle subfigure of 
Figure \ref{Fig:VERITASLightCurve}.
The red line provides the average flux reported in the 3FGL catalog.
The 3FGL catalog asserted that the RGB J2243+203 GeV flux did not exhibit significant variability when averaged month by month.
Figure \ref{Fig:FermiLongTermLightCurve} shows the 6-month light curve with a binning of 7 days.
The time-averaged flux over four years is shown in a solid red line.
The highest flux recorded within this six months is over a two-week period that includes 
both 17 and 24 December 2014 UTC, which coincides with the VERITAS detection.

\subsection{Swift X-Ray Telescope (XRT) observations}\label{SwiftAnalysis}
The VERITAS discovery of TeV emission from RGB J2243+203 triggered pointed observations with the \textit{Swift} space telescope \citep{Gehrels2004}.
The focusing X-ray telescope (XRT) on board \textit{Swift} is sensitive to X-ray photons in the energy range from 0.3 keV to 10 keV.
\textit{Swift} observed the source five times between MJD 57017.945 (2014-12-26 22:40:48.000 UTC) and 
57021.735 (2014-12-30 17:38:24.000 UTC).
The \textit{Swift} data archive also contains previous observations of the source taken between MJD 54956.2 and 56842.811.
Photon Counting (PC) readout mode was used to carry out the observations, and there is no photon pileup in the data.
The data were calibrated and cleaned with standard criteria with the xrtpipeline task using 
the calibration files as available in the Swift CALDB version 20140120.
Spectral analysis was performed with events recorded within a circle of 20 pixels, corresponding to $\sim 45^{\prime \prime}$
radius centered on the source position.
A nearby circular region of 40-pixel radius was used to estimate the background. 
The most recent response matrices (v.014) available in the Swift CALDB were used.

Table \ref{Tab:Swift} summarizes the \textit{Swift}-XRT observations of RGB J2243+203.
For the archival data and triggered observations, the X-ray energy spectrum between 2 keV and 10 keV was fitted with a power-law.
The Table shows the spectral index for the best-fit energy spectrum, and the integrated flux between 2 keV and 10 keV.
To ensure a reasonable constraint on the measured spectra, spectral energy distribution fitting was performed with bins that have at least of 20 counts in each bin.
However, the observations on MJD 57019.013 and MJD 5720.205 do not have enough bins with more than 20 counts to perform a spectral fit.
The cumulative data set of all five VERITAS-triggered observations can also be fitted with a power-law with a spectral index of $2.7 \pm 0.2$, 
and the integrated photon flux in the same energy band is $(3.6 \pm 0.6) \times 10^{-13} ~\textrm{cm}^{-�2} \textrm{s}^{-1}$.

The X-ray flux for the first archival observation is larger than all the other flux measurements, including the triggered observations, clearly representing an X-ray flux elevated state.  
The average flux of the archival observations and the triggered observations without the first measurement is  
$(3.5 \pm 1.2) \times 10^{-13} ~\textrm{cm}^{-2} \textrm{s}^{-1}$.
The light curve of the triggered observations is shown in the bottom subfigure of Figure \ref{Fig:VERITASLightCurve}.
The red line shows the average flux.
\textit{Swift}-XRT observations on 26 December 2014 show a mildly elevated flux, but the following nights did not show 
strong X-ray activity.

\begin{table}[h]
\centering
\begin{tabular}{ | c | c | c | c | c |}
  \hline
  Observation Start & Live Time & counts per second & Spectral Index & Integrated Flux (2 - 10 keV) \\
  MJD	              & seconds    &        $\times 10^{-2}$          &  & $\times 10^{-13}$  cm$^{-2}$s$^{-1}$\\
  \hline
    \multicolumn{4}{c}{Archival Observations}\\
  \hline
54956.200 		& 5414.137 &	$29.0 \pm 0.7$		& $2.62 \pm 0.06$ 		&  $26.4 \pm 1.5 $\\
56039.338 		& 1066.342 &	$6.9 \pm 0.8$		& $2.9 \pm 0.3 $ 		& 	$4.3 \pm 1.2 $\\
56460.179 		& 1028.892 &	$2.1 \pm 0.5$		& $2.5_{-0.6}^{+0.7} $ 	&	$3.6  \pm 2.0 $\\
56842.811 		& 1997.831 &	$3.9 \pm 0.4$		& $2.8 \pm 0.3 $ 		&	$2.8 \pm 0.7 $\\
  \hline
  \multicolumn{4}{c}{VERITAS Triggered Observations}\\
  \hline
57017.945  		& 1043.866 &	$8.1 \pm 0.1$		& $2.7 \pm 0.3$  		&	$6.6 \pm 1.5$\\
57018.944 		& 1001.413 &	$3.7 \pm 0.6$		& $3.2^{+0.6}_{-0.5} $ 	&	$1.5 \pm 0.6$\\
57019.013 		& 1418.466 &	$3.6 \pm 0.5$		& - 					&  - \\
57020.205 		& 1051.358 &	$4.0 \pm 0.6$		& -  				&  - \\
57021.735 		& 874.052   &	$4.4 \pm 0.7$		& $3.1 \pm 0.6$ 		& 	$2.5 \pm 0.9$ \\
  \hline  
\end{tabular}

\caption{Summary of the \textit{Swift}-XRT observations. 
For each day the X-ray energy spectrum between 2 and 10 keV can be fitted with a power-law.
The third column shows the spectral index of the best fit, and the fourth column shows the integrated flux between 2 and 10 keV. 
Spectral fitting failed for the observations on MJD 57019.013 and MJD 57020.205.}
\label{Tab:Swift}
\end{table}

\section{ Redshift Upper Limit}\label{Redshift}
An upper limit on the redshift of RGB J2243+203 can be derived using the \textit{Fermi}-LAT-measured energy spectrum 
and the VERITAS-measured energy spectrum.
This method was proposed by \cite{Georganopoulos2010} and has been applied by others to derive redshift upper limits (e.g. \cite{Aleksic} and \cite{HESSPKS}).
This method does not assume any source emission models. 
The only assumption is the observed VERITAS high-energy component of the energy spectrum is softer than the extrapolation of the \textit{Fermi}-LAT GeV energy spectrum. 
At a given energy within the VERITAS observed energy range,

\begin{equation}
 F(E)_{obs} < F(E)_{ext},
\end{equation}

\noindent where $F(E)_{obs}$ is the observed VHE flux at energy E, 
and $F(E)_{ext}$ is the flux derived at energy E from the extrapolation of the GeV energy spectrum measured quasi-simultaneous with VERITAS.
Using the above assumption, Equation 1 in \cite{Georganopoulos2010} gives the upper limit to the EBL-induced pair absorption optical depth at a given energy $\tau_{max}(E)$,
\begin{equation}
 \tau_{max}(E) = \ln(F(E)_{ext} / F(E)_{obs}).
\end{equation}
Similar to \cite{Aleksic} and \cite{HESSPKS}, this equation can be modified to include a correction term for the error on the observed VERITAS flux,
\begin{equation}
 \tau_{max}(E) = \ln \left( \frac{F(E)_{ext}}{F(E)_{obs} - 1.64 \times \Delta F(E)_{obs}} \right),
\label{Eq:OptDepth}
\end{equation}
where $\Delta F(E)_{obs}$ accounts for the error in the VERITAS flux measurements. 
$\tau_{max}(E)$ gives the upper limit of $\tau(E)$ as a function of energy.
The upper limit of $\tau(E)$ can be translated into an upper limit on the redshift by comparing with EBL models.
Intrinsic hardening of the energy spectrum at energies greater than the \textit{Fermi}-LAT component of the energy spectrum would increase the derived upper limit.
Although some hadronic models predict a spectral hardening at high energies (e.g.  \cite{CerrutiHadron}), in general, spectral hardening at TeV energies is not expected.
The derived redshift upper limit would be lower if there were intrinsic softening of the energy spectrum.
Softening of the energy spectrum at higher energies is possible, but a lower redshift would not contradict the derived upper limit.

\begin{figure}
  \centering
 \includegraphics[width=0.75\textwidth]{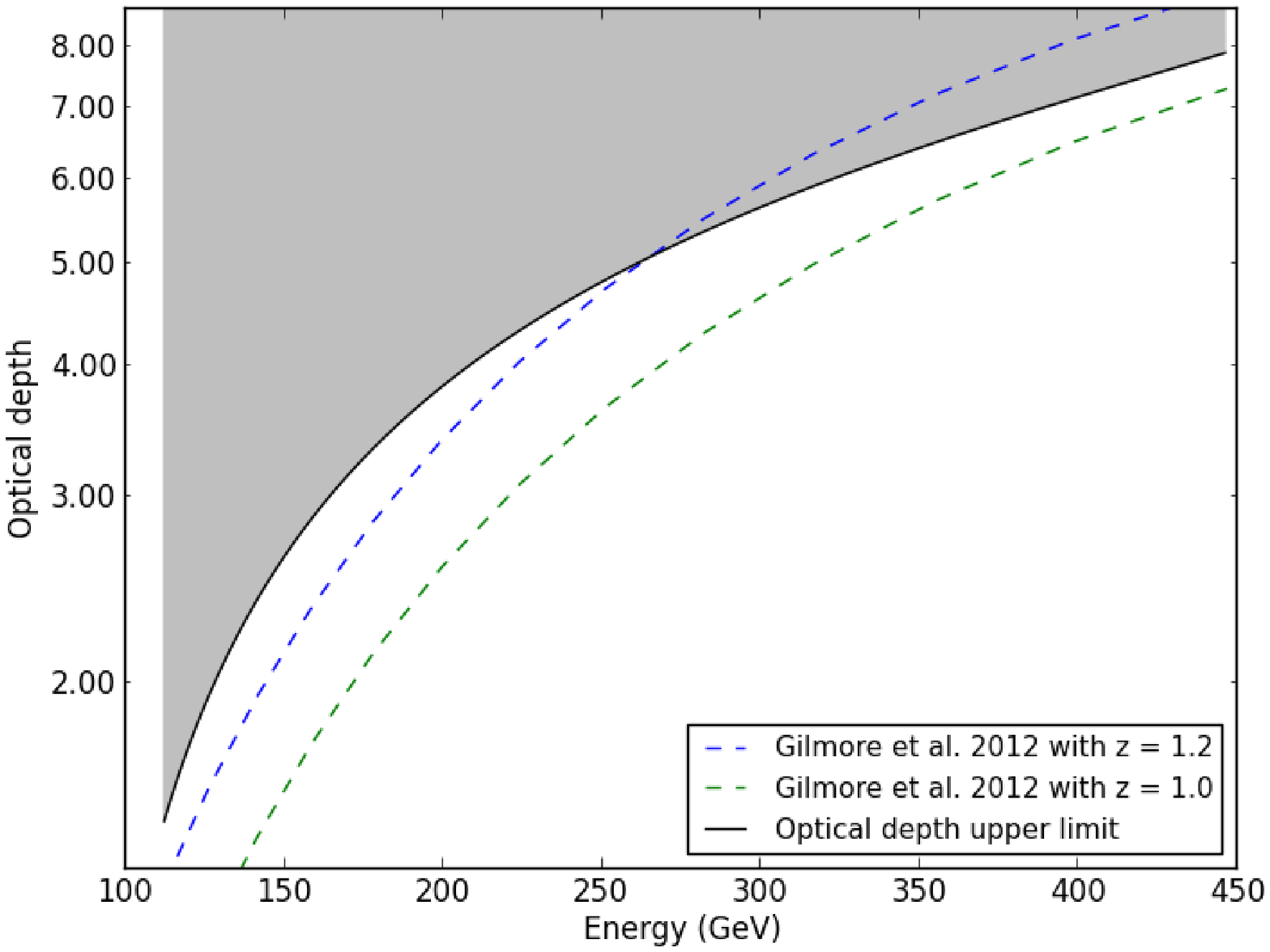}
 \caption{The black solid line indicates the maximum EBL-induced pair absorption optical depth at different energies along the line of sight to RGB J2243+203 that is derived from VERITAS and Fermi-LAT data.
 Two dashed lines indicate the modeled \citep{Gilmore2012} EBL-induced pair absorption optical depth as a function of energy for redshifts of 1.2 and 1.0.
 The shaded area is the excluded region.}
 \label{fig:TauMax}
\end{figure}

In Figure \ref{fig:TauMax} the upper limits of $\tau$ as a function of energy ($\tau_{max} \left( E \right)$) is shown by a solid black line.
The region marked in ash color is the excluded region.
The $\tau(E, z)$ distributions derived with the EBL model proposed by \cite{Gilmore2012} at redshifts of 1.2 and 1.0 are shown in dotted lines.

\begin{figure}
 \centering
 \includegraphics[width=0.75\textwidth]{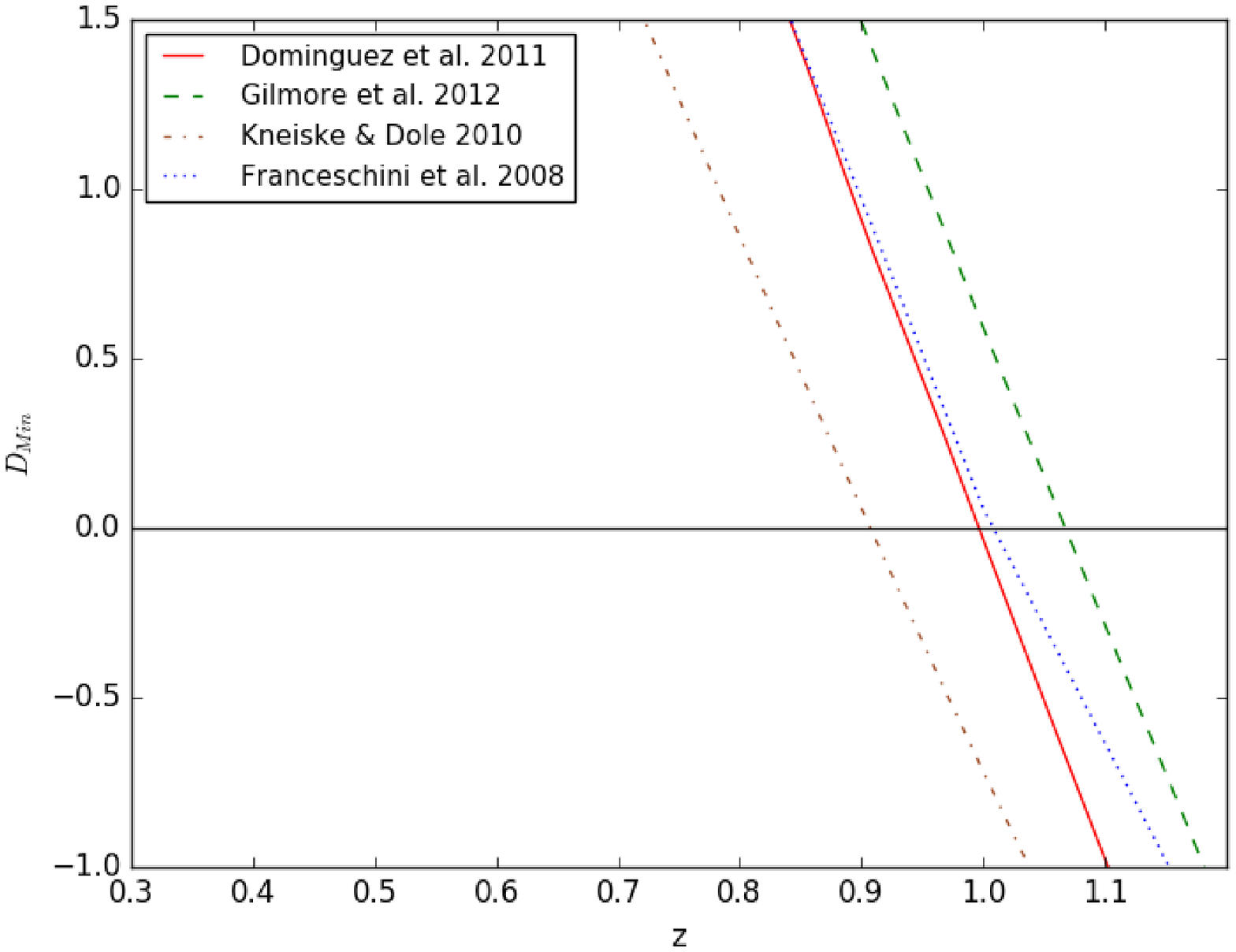}
 \caption{The minimum difference between the calculated maximum EBL-induced pair absorption optical depth and the model predicted EBL-induced pair absorption optical depth as a function of redshift. 
}
  \label{Fig:Combined}
\end{figure}

The minimum differences between $\tau(E, z)$ and $\tau_{max}$ ($D_{\textrm{min}}(z) = \tau(z,E) - \tau_{max}(E)$), over all energy bins of the VERITAS spectrum, for three different state of the art EBL models are shown in Figure \ref{Fig:Combined}.

At the 95\% confidence interval upper limit of the redshift $D_{\textrm{min}}$ becomes zero. 
As shown in Figure \ref{Fig:Combined}, the \cite{Gilmore2012}, \cite{Dominguez}, \cite{Kneiske}, and \cite{Franceschini2008} models predict upper limits of $z < $ 1.1, $z < $ 1.0, $z < 0.9$ and $z < $ 1.0, respectively.

\section{Discussion}\label{Sec:Discussion}

The simultaneous gamma-ray SED with \textit{Fermi}-LAT and VERITAS is shown in Figure \ref{Fig:MWLSED}.
The VERITAS flux points are not corrected for EBL absorption.
No other instruments observed RGB J2243+203 simultaneously with VERITAS.  
Observations by \textit{Swift}-XRT in X-rays were triggered after the discovery of gamma-ray activity.
The observed X-ray flux on 26 December 2014 shows a marginally elevated flux, 
but archival data shows the following nights do not have elevated fluxes.
While it is possible that the gamma-ray flare from RGB J2243+203 was not associated with a flare at longer wavelengths (i.e. an orphan gamma-ray flare), it is more likely than any activity at longer wavelengths was missed, and that the \textit{Swift}-XRT observations taken 67 hours after the VERITAS flare were measuring a post-flare state. 
It is well known that blazar emission models, even in the simplest one-zone synchrotron self-Compton scenario, are degenerate in the absence of simultaneous measurements of the synchrotron and inverse-Compton peaks \citep{Tavecchio98, Bottcher12, Cerruti13}.  For this reason, no specific modeling of the SED was performed.

\section{Conclusion}
VHE gamma-ray emission has been detected from RGB J2243+203 by the VERITAS observatory.
Earlier VERITAS upper limits from this source provide evidence that the source is variable in the VHE band, and suggest that the detection was obtained during an active period of the source.
The gamma-ray energy spectrum above 112 GeV can be fitted with a power-law with a spectral index of $4.6 \pm 0.5$, 
and a flux normalization at 0.15 TeV of $(0.63 \pm 0.11) \times 10^{-9} ~\textrm{cm}^{-2} \textrm{s}^{-1} \textrm{TeV}^{-1}$.
The gamma-ray energy spectrum between 1 GeV and 100 GeV measured by \textit{Fermi}-LAT quasi-simultaneous with VERITAS observations can be fitted with a power-law with a spectral index of $1.8 \pm 0.2$, and a flux normalization at 2.2 GeV of $(8.4 \pm 1.7) \times 10^{-12} ~\textrm{cm}^{-2} \textrm{s}^{-1} \textrm{MeV}^{-1}$. 
The integrated photon flux averaged over the time window is 
$(4.1 \pm 0.8) \times 10^{-8} ~\textrm{cm}^{-2} \textrm{s}^{-1}$, which is an order of magnitude larger than the 
four year averaged flux in the same energy range, $(4.0 \pm 0.1) \times 10^{-9} ~\textrm{cm}^{-2} \textrm{s}^{-1}$.
This supports the conclusion that the source was also active in the GeV energy band during the time of VHE detection.
The VERITAS discovery of TeV emission from RGB J2243+203 triggered observations with the \textit{Swift}-XRT.
However, the first \textit{Swift}-XRT observation was obtained 67 hours after the VERITAS observations.
This measurement indicates a mildly elevated state in flux, 
but the following nights have the same level of flux as archival data.
While \cite{Meisner2010} placed a lower limit for the redshift of $\textrm{z} > 0.39$, 
this article places EBL-model-dependent upper limits for the redshift, in the range from $z~<~0.9$ to $z~<~1.1$, depending on the EBL model used.

VERITAS is supported by grants from the U.S. Department of Energy Office of Science, the U.S. National Science Foundation and the Smithsonian Institution, and by NSERC in Canada. We acknowledge the excellent work of the technical support staff at the Fred Lawrence Whipple Observatory and at the collaborating institutions in the construction and operation of the instrument. The VERITAS Collaboration is grateful to Trevor Weekes for his seminal contributions and leadership in the field of VHE gamma-ray astrophysics, and for his interest in the wider applications of IACTs, which made this study possible.

\end{document}